\documentstyle[aps,epsf,multicol,amsfonts,graphicx,bbm]{revtex}

\begin{document}

\draft

\title{Time symmetry in Rigged Hilbert Spaces}
\author{Kim Bostr\"om}
\address{Institut f\"ur Physik, Universit\"at Potsdam, 14469 Potsdam, Germany}

\author{}
\address{
}
\date{\today}
\maketitle

\begin{abstract}

Despite the fact that the fundamental physical laws are symmetric in time, most observed processes do not show this symmetry. Especially the phenomenon of decay seems to involve a kind of irreversibility that makes the definition of a microscopic arrow of time possible. Such an intrinsic irreversibility is incorporated within the Rigged Hilbert Space quantum mechanics of the Brussels School, contrasting to the statements of standard quantum mechanics. As shown in this paper, the formalism bears significant advantages in the description of decaying systems, however the breaking of time symmetry can be avoided.

\end{abstract}

\begin{multicols}{2}
\narrowtext

\section{Introduction}

\begin{figure}
	\[\includegraphics[width=0.4\textwidth]{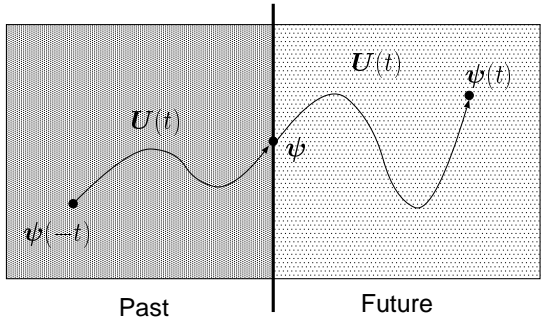}\]
	\vspace*{-0.5cm}\caption{\small In standard quantum mechanics every state of a closed system can at the same time be regarded as an initial state and a final state.}\label{fig:evol1}
\end{figure}

Asking for the time evolution of a system means solving the Schr\"odinger equation $i\hbar\frac\partial{\partial t}\psi=H\psi$. 
Restricted to the case of closed systems - which is done from now on - this equation can formally be solved by
\begin{equation}\label{prop1}
	\psi(t)=e^{-iHt}\psi\quad,
\end{equation}
with $\hbar=1$ for the sake of simplicity and $\psi=\psi(0)$. 
The whole dynamics is thus governed by the unitary propagator 
$U(t)=e^{-iHt}$ which maps any given ``initial state'' $\psi$ at time $t=0$ to its propagated counterpart $\psi(t)$ at time $t$.  Quantum mechanical determinism is expressed by the fact that $\psi$ can also be regarded as a ``final state'' into which a state $\psi(-t)$ evolves by satisfying $U(t)\psi(-t)=\psi$ (see fig.~\ref{fig:evol1}). 
Since equation~(\ref{prop1}) is valid for all times $t\in\mathbbm R$, 
giving $\psi(-t)=U(-t)\psi$, 
one is led to the operator identity
\begin{equation}\label{group1}
	U(-t)U(t)={\mathbbm 1}\quad.
\end{equation}
In other words: The set of evolution operators $\{U(t)\}$ forms a unitary group $\mathcal U$ with the group property~(\ref{group1}), giving any propagator $U(t)$ its inverse element $U(-t)=U^\dagger(t)$. The physical meaning of this mathematical statement is profound: in quantum mechanics there is no microscopic arrow of time. Any state $\psi\in\mathcal H$ can be propagated to the future as well as to the past, so that there is a complete time symmetry.

In order to explain an arrow of time despite these facts, there are several ideas. A few of them are:

\begin{enumerate}
\item
	Any measurement on the system interrupts the deterministic evolution of the state by applying random projections (plus normalization) instead of unitary operations (the ``collapse of the wave function''). This process cannot be reversed.
\item
	Real physical systems can never be regarded as closed. There is always an interacting environment with many unobservable degrees of freedom, inducing non-unitary time evolution of the observed system.
\item
	Time symmetry in classical mechanics is only broken with respect to macroscopic properties of a system (e.g. temperature or entropy). Considering macroscopic observables in quantum mechanics, this effect also shows up (coarse graining argument).
\end{enumerate}

All these ideas have been worked out to elaborate physical models with strong experimental support and all of them break time symmetry (see e.g.~\cite{zeh,giulini}). So one could be content with the situation and regard those models as fairly good explanations of an arrow of time, which is indeed an undeniable part of human experience.

A group of scientists around \emph{Ilja Prigogine} (nobel prize in chemistry 1977), referred to as the ``Brussels School'', is not satisfied with the situation and has worked out a quantum mechanical formalism especially applying to unstable systems  (i.e. systems involving decay), where time symmetry is claimed to be broken at the microscopic level (see e.g.~\cite{petrosky1991,antoniou1993,antoniou1993b,bohm1998,doebner,bohm1989,likhoded1997}). 
\begin{figure}
	\[\includegraphics[width=0.4\textwidth]{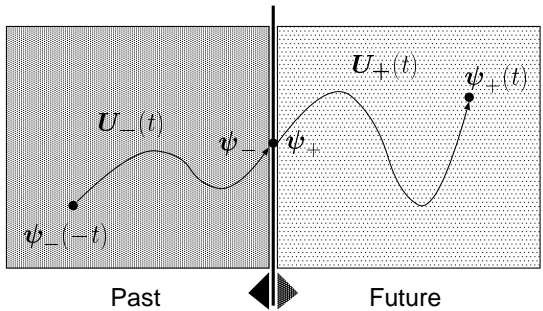}\]
	\vspace*{-0.5cm}\caption{\small In Brussels School's quantum mechanics every state is whether an initial or a final state. Time evolution falls apart into two different directions.}\label{fig:evol2}
\end{figure}
This fact is expressed by the splitting of the unitary group $\mathcal U$ into two semigroups $\mathcal U_-$
and $\mathcal U_+$ 
of propagators for negative and positive times, such that past and future states can no longer be regarded as equivalent counterparts. 
This leads to a so-called ``intrinsic irreversibility'', implying a microscopic arrow of time (see fig.~\ref{fig:evol2}). 
\smallskip

This paper is separated into two different parts. The first part is dedicated to a review of the Rigged Hilbert Space formalism worked out by the Brussels School. It is shown how it works, how it applies to physical systems and what its advantages over standard quantum mechanics are. The second part is devoted to the role of time symmetry within the Rigged Hilbert Space formalism. Apart from its practical advantages, the Brussels School points out that a microscopic arrow of time can be implemented here. It will be shown that such an implementation is possible but not mathematically necessary.

\section{The basic concepts}

\subsection{Spectral decompositions}

Writing down the formal solution~(\ref{prop1}) 
of a closed system's Schr\"odinger equation is not yet really satisfying. How can expectation values of observables be numerically calculated? 
To this aim one has to find a spectral decomposition of the system's Hamiltonian $H$, thus one has to solve the time-independent Schr\"odinger equation $H\psi=E\psi$.
Since $H$ is a self-adjoint operator, its spectral decomposition is complete and orthogonal and may therefore be used as a spectral decomposition of the unity operator $\mathbbm 1$. Furthermore the spectrum of $H$ is real and consits in general of a discrete part $\sigma_d$ and a continuous part $\sigma_c$. 
In special cases there are other parts, e.g. singular continuous and dense spectra, but these cases are excluded here. With this restriction, using the Dirac bra-ket notation, the spectral decomposition of the unity operator reads 
in general
\begin{equation}
	{\mathbbm 1}=\sum_n |E_n\rangle\langle E_n|
	+\int_{\sigma_c} dE\,\rho(E)\,|E\rangle\langle E|\quad,
\end{equation}
where $H|E_n\rangle=E_n|E_n\rangle$, $H|E\rangle=E|E\rangle$ and $\rho(E)$ is the energy density.
As soon as the energy decomposition of the initial state $|\psi\rangle$, generally given by
\begin{equation}
	|\psi\rangle=\sum_n \psi_n|E_n\rangle+\int_{\sigma_c} dE\,\rho(E)\,\psi(E)
	|E\rangle\quad,
\end{equation}
is known, the propagated state is also known and reads
\begin{eqnarray}
	|\psi(t)\rangle&=&\sum_n \psi_n e^{-iE_n t}|E_n\rangle\nonumber \\
	&&+\int_{\sigma_c} dE\,\rho(E)\,\psi(E)e^{-iEt}|E\rangle\quad.
\end{eqnarray}
The expectation value of an observable $A$ can be calculated for any time $t$ using
\begin{equation}
	<A>(t)=\langle\psi(t)|A|\psi(t)\rangle\quad.
\end{equation}
So the fundamental key to the dynamics of a system is the spectral decomposition of its Hamiltonian. All dynamical properties are monitored by this decomposition and here is where the new formalism enters. There is a class of systems where the Brussels Schools's spectral decomposition of the Hamiltonian differs from the usual decomposition: systems envolving decay, called \emph{unstable systems}. 

\subsection{Resolvent techniques}

\begin{figure}[h!]
	\[\includegraphics[width=0.5\textwidth]{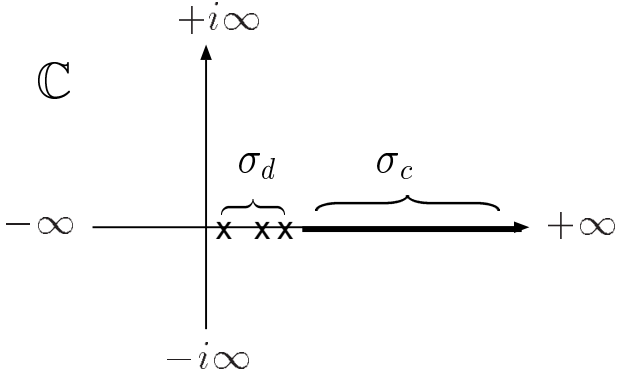}\]
	\vspace*{-0.5cm}\caption{\small The eigenvalues of the Hamiltonian $H$ are the irregularities of its resolvent $R(z)=\frac1{z-H}$. Since $H$ is self-adjoint, they all take place on the real axis, where the discrete eigenvalues $E_n\in\sigma_d$ are singularities and the continuous eigenvalues $E\in\sigma_c$ form a cut in the complex plane.}\label{fig:spectrum}
\end{figure}

The basic concept of the Brussels School's formalism is to find a different spectral decomposition of the Hamiltonian. In contrast to the method of \emph{complex scaling} (see~\cite{balslev}), the Hamiltonian is not modified and remains self-adjoint; it only acquires a different representation involving generalized eigenvectors contained in an extended distribution space. The modified decomposition can most easily be achieved by using the Hamiltonian's \emph{resolvent}
\begin{equation}
	R(z):=\frac1{z-H}\quad.
\end{equation}
The spectrum of $H$ is defined by the irregularities of its resolvent. The discrete eigenvalues are the singularities of $R(z)$, i.e. those points in the complex plane where the operator $R$ is not everywhere defined on the Hilbert space $\mathcal H$. The continuous eigenvalues are those points where $R$ is an unbound (and thus non-continuous) operator on $\mathcal H$. Anywhere else on the complex plane the resolvent is an analytic function mapping complex numbers onto the set of linear bounded operators. As $H$ is self-adjoint, any irregularities of $R$ take place on the real axis, where the discrete eigenvalues form the point spectrum and the continuous eigenvalues define a cut in the complex plane (see fig.~\ref{fig:spectrum}). 
It is possible to find an analytic continuation of $R$ from both sides of the cut, each leading to a different operator
\begin{equation}
	R_\pm(E):=\frac1{E-H\pm i\epsilon}=R_\mp^\dagger(E),
\end{equation}
with $\epsilon\rightarrow+0$ understood here and in the following. Note that this continuation is not possible at singularities embedded in the continuous spectrum. Note further that on the real axis outside the continuous spectrum both operators $R_+$ and $R_-$ coincide with $R$. The two continued resolvents are known as the \emph{retarded} and the \emph{advanced} \emph{Green operator}, respectively.

The resolvent can be used to construct eigenvectors. 
The residuum of the resolvent at a discrete eigenvalue yields the dyadic product of the corresponding eigenvector:
\begin{equation}
	\lim_{z\rightarrow E_n\pm i\epsilon}(z-E_n)R(z)
	=|E_n^\pm\rangle\langle E_n^\pm|\quad.
\end{equation}
Note that both dyadic products project onto the same subspace corresponding to $E_n$. If the discrete eigenvalue is isolated, i.e. not embedded in the continuous spectrum, both limits from above and below the real axis coincide.  In the following we will assume that all discrete eigenvalues are isolated and we therefore have
\begin{equation}
	\lim_{z\rightarrow E_n}(z-E_n)R(z)
	=|E_n\rangle\langle E_n|\quad.
\end{equation}
A discrete eigenvector $|E_n\rangle$ of $H$ can be obtained by applying the residual resolvent to any Hilbert vector $|\psi\rangle$ being not orthogonal to the eigenspace of $E_n$. :
\begin{equation}\label{eigenvector1}
	|E_n\rangle=\frac1{\psi_n}\lim_{z\rightarrow E_n}(z-E_n)R(z)
	|\psi\rangle\quad,
\end{equation}
where $\psi_n=\langle E_n|\psi\rangle\neq0$ is an adequate normalization constant, such that $\langle E_n|E_m\rangle=\delta_{nm}$.

In analogy to the discrete case the residuum of the resolvent at one of the continuous eigenvalues yields the dyadic product of the corresponding continuous eigenvector: 
\begin{equation}\label{dyad1}
	\lim_{z\rightarrow E\pm i\epsilon}(z-E)R(z)
	=|E^\pm\rangle\langle E^\pm|\quad.
\end{equation}
Note that both dyadic products are equivalent representations of the $\delta$-Operator $\delta(E-H_c)$, defined by
\begin{equation}
	\int_{\sigma_c} dE\,\rho(E)\,\delta(E-H_c)f(E):=f(H_c)\quad,
\end{equation}
where $H_c$ is the continuous part of the total Hamiltonian $H$.
Since the continuous eigenvalues are not isolated, there are always pairs of eigenvectors. They correspond to retarded and advanced solutions of the continuous eigenvalue equation and are known as \emph{incoming} and \emph{outgoing waves}. Both solutions can equivalently be used in the spectral decomposition and they can be constructed by
\begin{equation}\label{eigenvector2}
	|E^\pm\rangle=\frac1{\psi_\pm(E)}
	\frac{\pm i\epsilon}{E-H\pm i\epsilon}|\psi\rangle\quad,
\end{equation}
where $\psi_\pm(E)=\langle E^\pm|\psi\rangle\neq0$ is a suitable normalization constant, such that $\langle E^\pm|{E'}^\pm\rangle=\delta(E-E')$.

Using dyadic products, the retarded and advanced Green operators can be written as
\begin{equation}
	E\in\sigma_c:\quad R_\pm(E)=\frac{{\mathcal P}_E}{E-H}\mp i\pi|E\rangle\langle E|\quad,
\end{equation}
where $|E\rangle\langle E|$ is any of the two dyadic products $|E^\pm\rangle\langle E^\pm|$ and the principal value distribution ${\mathcal P}_E$  is defined by
\begin{equation}
	\int dE'\,{\mathcal P}_E\,f(E')
	:=\int_{E-\epsilon}^{E+\epsilon}dE'\,f(E')\quad.
\end{equation}
The difference of the two Green operators gives the dyadic product of the continuous eigenvector:
\begin{equation}\label{dyad2}
	\frac{i}{2\pi}\big\{R_+(E)-R_-(E)\big\}=|E^\pm\rangle\langle E^\pm|\quad.
\end{equation}
Altogether, the spectral decomposition of the unity operator may be expressed by the exclusive use of the resolvent:
\begin{eqnarray}\label{unity}
	{\mathbbm 1}&=&\sum_n \lim_{z\rightarrow E_n}(z-E_n)R(z)\nonumber \\
	&&+\frac{i}{2\pi}\int_{\sigma_c}dE\,\rho(E)\,\big\{R_+(E)-R_-(E)\big\}\quad.
\end{eqnarray}
This unity decomposition will be used as the starting point of the Brussels Schools's unity spectral decomposition.

\subsection{The second Riemann sheet}

\begin{figure}[h!]
	\[\includegraphics[width=0.45\textwidth]{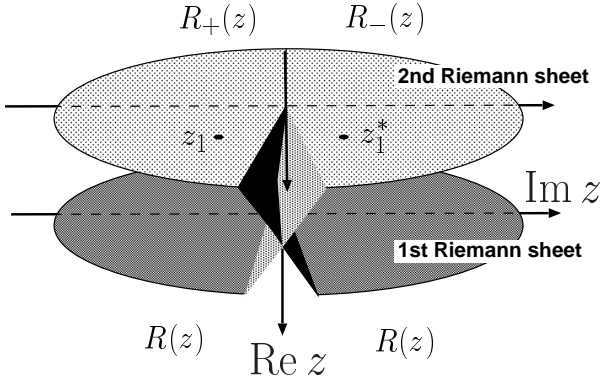}\]
	\vspace*{-0.5cm}\caption{\small The second Riemann sheet. New singularities of the analytically continued resolvent may appear, both complex conjugated to another.}\label{fig:riemann}
\end{figure}

The resolvent can be analytically continued to the complex plane beyond the real axis entering the \emph{second Riemann sheet}:
\begin{equation}
	R_\pm(z):=\lim_{E\rightarrow z} R_\pm(E),\quad E\in\sigma_c\quad.
\end{equation}
On the one half of the complex plane where the continuation started (the first Riemann sheet), both operators coincide with the original resolvent. On the other half (the second Riemann sheet) it might happen that new singularities appear, and that in a pairwise manner: If $z_k$ is a singularity of $R_+(z)$ in the lower half of the complex plane, then $z_k^*$ is a singularity of $R_-(z)$ in the upper half (see fig.~\ref{fig:riemann}). 
Unstable systems are exactly those with a resolvent having complex poles on the second Riemann sheet.
The second Riemann sheet poles act as complex resonances and cause the decay of certain states.

\subsection{An extended spectral decomposition}\label{deform}

Consider the continuous part of the unitary decomposition~(\ref{unity}):
\begin{equation}\label{unity_c}
	{\mathbbm 1}_c:=\frac{i}{2\pi}\int_{\sigma_c}dE\,\rho(E)\,
	\big\{R_+(E)-R_-(E)\big\}\quad.
\end{equation}
The following manipulations of ${\mathbbm1}_c$ are reducing its domain. As a consequence one is forced to use a smaller \emph{test space}, enlarging at the same time the distribution space and inducing this way a certain \emph{Gelfand tripel} or \emph{Rigged Hilbert Space} where the modified completeness relation remains valid. We delay this discussion to section~\ref{rhs}.

\begin{figure}[h!]
	\[\includegraphics[width=0.5\textwidth]{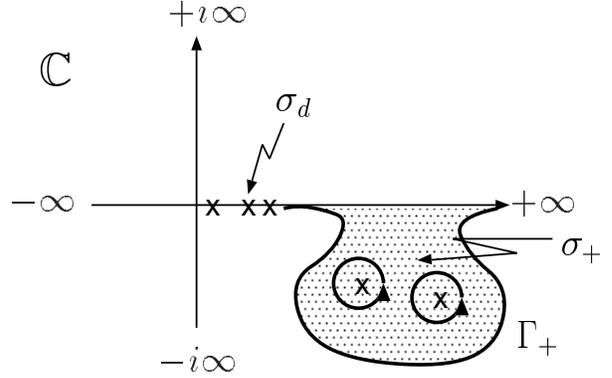}\]
	\vspace*{-0.5cm}\caption{\small Analytically deforming the integration path along $\sigma_c$ to a path $\Gamma_+$ one crosses the singularities of $R_+(z)$, which have to be split off from the integration.}\label{fig:spectrum2}
\end{figure}

If one analytically deforms the integration path along the continuous spectrum $\sigma_c$ to a path $\Gamma_+$ leading through the lower half of the complex plane, one enters the second Riemann sheet and crosses the singularities of $R_+(z)$. Let the set of these singularities be defined by $\sigma_+=\{z_k\}$ and be enclosed by $\sigma_c\cup\Gamma_+$. In the course of the deformation circular integrations around the singularities are split off, each leading to a residuum of $R_+(z)$ at these points (see fig.~\ref{fig:spectrum2}).
The modified operator now reads
\begin{eqnarray}\label{unity_c2}
	{\mathbbm 1}_c^+&=&\sum_k\lim_{E\rightarrow z_k}(E-z_k)R_+(E)\nonumber \\
	&&+\frac{i}{2\pi}\int_{\Gamma_+}dz\,\rho(z)\,
	\big\{R_+(z)-R_-(z)\big\}\quad.
\end{eqnarray}
Introducing a \emph{complex distribution} $D_\pm(E)$, defined by
\begin{equation}\label{compl_dist}
	\int_{\sigma_c} dE\,D_\pm(E)\,f(E):=\int_{\Gamma_\pm}dz\,f(z)\quad,
\end{equation}
where $\Gamma_-=\Gamma_+^*$ and $f(z)$ being the analytic continuation of $f$ to respectively the upper and lower half of the complex plane, one can rewrite ${\mathbbm1}_c^+$ as
\begin{eqnarray}
	{\mathbbm 1}_c^+&=&\sum_k\lim_{E\rightarrow z_k}(E-z_k)R_+(E)\nonumber \\
	&&+\frac{i}{2\pi}\int_{\sigma_c}dE\,\rho(E)\,D_+(E)
	\big\{R_+(E)-R_-(E)\big\}\quad.
\end{eqnarray}
In analogy to the standard case one defines dyadic products of \emph{generalized eigenvectors} of $H$ by
\begin{equation}
	|f_k^+\rangle\langle f_k^-|
	:=\lim_{E\rightarrow z_k}(E-z_k)R_+(E)\quad
\end{equation}
for the discrete case and, using~(\ref{dyad2}),
\begin{equation}
	|f^+(E)\rangle\langle f^-(E)|
	=D_+(E)|E\rangle\langle E|\quad
\end{equation}
for the continuous case, where $|E\rangle\langle E|$ is any of the two dyadic products $|E^\pm\rangle\langle E^\pm|$.
Hence we have
\begin{eqnarray}
	\int_{\sigma_c}dE\,\rho(E)\,|E\rangle\langle E|
	&=&\sum_k|f_k^+\rangle\langle f_k^-|\nonumber \\
	&+&\int_{\sigma_c}dE\,\rho(E)\,|f^+(E)\rangle\langle f^-(E)|
	\label{unity_2rep}.
\end{eqnarray}
The generalized eigenvectors appear as pairs of right and left eigenvectors. The discrete eigenvectors are called \emph{Gamov vectors} and can be constructed by:
\begin{eqnarray}\label{g_eigenvector1a}
	|f_k^+\rangle&=&\frac1{\psi_k}\lim_{E\rightarrow z_k}(E-z_k)R_+(E)
	|\psi\rangle,\\
	\langle f_k^-|&=&\frac1{\psi_k}
	\lim_{E\rightarrow z_k}(E-z_k)\langle\psi|R_+(E)\label{g_eigenvector1b}
	,
\end{eqnarray}
where $|\psi\rangle$ is a suitable vector and $\psi_k=\langle f_k^+|\psi\rangle\neq0$ is a normalization constant, such that $\langle  f_k^-|f_l^+\rangle=\delta_{kl}$.
By adjunction one sees that $|f_k^+\rangle\neq| f_k^-\rangle$. Furthermore we have $\langle f_k^\pm|f_k^\pm\rangle=\|f_k^\pm\|=0$. These objects are left and right eigenvectors of the Hamiltonian $H$ corresponding to complex eigenvalues:
\begin{eqnarray}
	H|f_k^+\rangle&=&z_k|f_k^+\rangle,\\
	\langle f_k^-|H&=&\langle f_k^-|z_k\quad.
\end{eqnarray} 
Since the Gamov vectors have zero norm, they all coincide in Hilbert space with the null vector, which is not allowed to be an eigenvector. Also, since $H$ is self-adjoint, it cannot have any eigenvectors corresponding to complex eigenvalues. Concluding, the Gamov vectors cannot be elements of the Hilbert space. In fact they are distributions in a suitably chosen distribution space (see section~\ref{intrinsic}).
The continuous eigenvectors can be constructed in two ways:
\begin{eqnarray}\label{g_eigenvector2}
	|f^\pm(E)\rangle=D_\pm(E)|E^+\rangle&\Leftrightarrow&
	\langle f^\pm(E)|=D_\mp\langle E^+|,\\
	|g^\pm(E)\rangle=D_\pm(E)|E^-\rangle&\Leftrightarrow&
	\langle g^\pm(E)|=D_\mp(E)\langle E^-|.
\end{eqnarray}
By application to test vectors with analytic extensions to the complex plane one sees that
\begin{eqnarray}
	D_+(E)|E^-\rangle&=&|E^-\rangle\\
	D_-(E)|E^+\rangle&=&|E^+\rangle\quad.
\end{eqnarray}
Hence there are two equivalent bases given by
\begin{eqnarray}
	B^+_1&=&\{|E_n\rangle\langle E_n|,|f^+_k\rangle\langle f^-_k|,|f^+(E)\rangle\langle E^+|\},\\
	B^+_2&=&\{|E_n\rangle\langle E_n|,|f^+_k\rangle\langle f^-_k|,|E^-\rangle\langle g^-(E)|\}.
\end{eqnarray}
Both bases can equally be used to span the Hilbert space $\mathcal H$. Arbitrarily we select $B_1^+$ and obtain a \emph{retarded} unity decomposition
\begin{eqnarray}
	{\mathbbm1}_+&=&\sum_n |E_n\rangle\langle E_n|
	+\sum_k |f_k^+\rangle\langle f_k^-|,\nonumber \\
	&&+\int_{\sigma_c} dE\,\rho(E)\,|f^+(E)\rangle\langle E^+|\quad.
	\label{unity_ret}
\end{eqnarray}
In complete analogy one can modify the spectral decomposition by deformation of the integration path to $\Gamma_-=\Gamma_+^*$ in the \emph{upper} half of the complex plane, entering the domain of $R_-(z)$ with singularities at $\sigma_-=\{z_k^*\}$. One then obtains the two bases
\begin{eqnarray}
	B^-_1&=&(B_1^+)^\dagger,\\
	B^-_2&=&(B_2^+)^\dagger,
\end{eqnarray}
such that
\begin{eqnarray}\label{g_eigenvector3}
	|f_k^-\rangle&=&\frac1{\psi_k}\lim_{E\rightarrow z_k^*}(E-z_k^*)R_-(E)
	|\psi\rangle\\
	\langle f_k^+|&=&\frac1{\psi_k}
	\lim_{E\rightarrow z_k^*}(E-z_k^*)\langle\psi|R_-(E)\\
	\quad|f^-(E)\rangle&=&|E^+\rangle,
	\quad\langle f^+(E)|=D_-(E)\langle E^+|,\\
	\quad|g^-(E)\rangle&=&D_-(E)|E^-\rangle,
	\quad\langle g^+(E)|=\langle E^-|\quad.
\end{eqnarray}
Arbitrarily choosing $B_1^-$ one obtains an \emph{advanced} unity decomposition:
\begin{eqnarray}
	{\mathbbm1}_-&=&\sum_n |E_n\rangle\langle E_n|
	+\sum_k |f_k^-\rangle\langle f_k^+|\nonumber \\
	&&+\int_{\sigma_c} dE\,\rho(E)\,|E^+\rangle\langle f^+(E)|\quad,
	\label{unity_ad}
\end{eqnarray}
which is just the adjoint of the retarded unity decomposition, so $({\mathbbm1}_+)^\dagger\equiv {\mathbbm1}_-$. Note, however, that both decompositions are equivalent representations of the unity operator, thus in the sense of an operator identity we have ${\mathbbm1}_+={\mathbbm1}_-={\mathbbm1}$.

\subsection{Using the extended decomposition}\label{advantages}

The unity decompositions obtained in the preceding section can be used for different representations of vectors and operators. Using the retarded decomposition the Hamiltonian reads
\begin{eqnarray}
	H&=&\sum_n E_n|E_n\rangle\langle E_n|
	+\sum_k z_k|f_k^+\rangle\langle f_k^-|\nonumber \\
	&&+\int_{\sigma_c} dE\,\rho(E)\,E|f^+(E)\rangle\langle E^+|\quad.
\end{eqnarray}
The adjoint of the above expression yields the Hamiltonian in the \emph{advanced} decomposition. Since both decompositions are equivalent representations, one sees that $H^\dagger=H$, i.e. the Hamiltonian remains self-adjoint.

The propagator in the retarded decomposition reads
\begin{eqnarray}
	U(t)&=&\sum_n e^{-iE_nt}|E_n\rangle\langle E_n|
	+\sum_k e^{-iz_kt}|f_k^+\rangle\langle f_k^-|\nonumber \\
	&&+\int_{\sigma_c} dE\,\rho(E)\,e^{-iEt}|f^+(E)\rangle\langle E^+|\quad.
\end{eqnarray}
Since the complex numbers $z_k$ are located in the lower half of the complex plane, the contribution from the Gamov vectors decays in the future. If an initial Hilbert state $|\psi\rangle$ has Gamov components, they will disappear for $t\rightarrow\infty$. So if the state has no rotating components, i.e. $\langle E_n|\psi\rangle=0$, the whole state will disappear in the future with the decay rates of the Gamov contributions given by the imaginary part of the contributing $z_k$:
\begin{equation}
	\psi_k^+ e^{-iz_kt}=\psi_k^+ e^{-\gamma_kt} e^{-i\nu_kt}\stackrel{t}{\longrightarrow}0\quad,
\end{equation} 
with $\psi_k^+=\langle f_k^+|\psi\rangle$, $z_k=\nu_k-i\gamma_k$ and $\nu_k,\gamma_k>0$.
Each contribution from a Gamov state also induces a certain energy shift, given by the real part of the $z_k$.
The advantage of the extended spectral decomposition is getting clear now: Not only the stable frequencies can immediately be read from the spectrum, but the decay rates and the energy shifts, too. By additionally checking the amplitudes $\psi_k^+$ between the initial state and the Gamov vectors one has immediate access to the dissipative dynamics of the system.

\subsection{A closer look}

Apart from the contribution of the Gamov vectors one has to consider the contribution from the generalized continuous eigenvectors. Let us call it the \emph{background contribution}. Since the decomposition involves the complex distribution $|f^+(E)\rangle=D_+(E)|E\rangle$, a test vector $\varphi$ has to be introduced. The background contribution then reads:
\begin{eqnarray}
	\langle\varphi|\psi_{bg}(t)\rangle&:=&\int_{\sigma_c}dE\,\rho(E)\,
	e^{-iEt}\langle\varphi|f^+(E)\rangle\langle E^+|\psi\rangle\\
	&=&\int_{\sigma_c}dE\,\rho(E)\,D_+(E)\psi_+(E)
	e^{-iEt}\varphi_+^*(E)\label{bg1}
\end{eqnarray}
where $\psi_+(E)=\langle E^+|\psi\rangle$ and $\varphi_+^*(E)=\langle\varphi|E^+\rangle$.
Due to $D_+(E)$ the integration path is transformed into the curve $\Gamma_+$ below the complex eigenvalues $z_k$. 
\begin{equation}
	\langle\varphi|\psi_{bg}(t)\rangle=\int_{\Gamma_+}dz\,\rho(z)\,e^{-izt}
	\psi(z)\varphi^*(z)\quad,\label{bg2}
\end{equation}
where $\psi_+(z)$ and $\varphi_+^*(z)$ are the analytic continuations of respectively $\psi_+(E)$ and $\varphi_+^*(E)$ to the lower half of the complex plane (provided their existence).
The curve $\Gamma_+$ is arbitrary as long as $\Gamma_+\cup\sigma_c$ surrounds $\sigma_+$, i.e., $\Gamma_+$ is placed \emph{below} the poles $z_k$, which all have negative imaginary parts. For $t>0$, due to the exponential factor $e^{-izt}$, the background contribution thus becomes more and more neglectable in comparison with the pole contributions. For initial states with high decay rates, i.e. pole contributions with small imaginary parts, the background contribution may be neglected already for small times $t>0$. This approximation is called the \emph{pole approximation} and it is essentially equivalent to the \emph{Wigner-Weisskopf approximation} (see~\cite{ww}). The time evolution of the initial state may be approximated by an exponential decay and its energy spectrum by a Breit-Wigner distribution. This approximation is very popular, e.g. in quantum optics, and indeed in most cases it is a very good one.

\subsection{Do divergences break time symmetry?}

Obviously the above argument does not hold for negative times. Here, the contributions from the Gamov vectors diverge. In several publications of the Brussels School (see e.g.~\cite{petrosky1991} or \emph{A. Bohm} in~\cite{doebner}) this is an argument in favour of a broken time symmetry, since for negative times the \emph{advanced} decomposition~(\ref{unity_ad}) is the better choice. However, as will be shown below, this is a rather intuitive reasoning and does not enforce a break of time symmetry. 

Let us see what happens with the retarded decomposition of a Hilbert state for negative times. Of course the Gamov contributions now diverge, but they are compensated by the background contribution. For $t\rightarrow-\infty$ the background integral~(\ref{bg2}) gains influence, since the integration path $\Gamma_+$ is placed \emph{below} the complex eigenvalues $z_k$. The same argument, that for increasing $t>0$ the background contribution can be neglected in comparison with the pole contributions, is turned upside down for $t<0$. The smaller imaginary part of $\Gamma_+$ induces a higher divergence than the contribution of the poles. By construction their divergences cancel each other, since the sum of both yields the contribution of the standard continuous eigenvectors (see eq.~(\ref{unity_2rep})), so a divergence of an initial Hilbert state backwards in time is prevented. The divergence of the Gamov contributions is even more than compensated: If a Hilbert state decays in the future, it will also decay in the past. 
To see this one can use the advanced decomposition~(\ref{unity_ad}) to investigate the system's evolution to the past and one would find a completely symmetric situation with majorizing pole contributions decaying backwards in time.

The situation is different if a Gamov vector $|f_k^+\rangle$ is taken as the initial state. Now there is only one pole contribution $z_k$ decaying in the future and diverging in the past. But Gamov vectors are only ``existing'' in the spectral representation of Hilbert states and they themselves do not refer to elements of physical reality, similiar to all other ``generalized eigenstates'' appearing in spectral decompositions. 
As constituted by the axiomatic foundations of quantum mechanics a physically realizable state is represented by a Hilbert vector (actually a family of Hilbert vectors normalized to unity and differing only in phase). Gamov states are no Hilbert vectors, so their properties are not expected to be physically meaningful.

Concluding, both decompositions are equivalent, but the use of the retarded decomposition is more \emph{intuitive} for $t>0$. However, this should not be taken as the reason for a broken time symmetry.
A more convincing argument in favour of a broken time symmetry is the splitting of \emph{test spaces}. As we have seen in~(\ref{bg2}), analytic continuations of the test functions are needed to have the decompositions well-defined. In section~\ref{rhs} we will get back to this point.

\subsection{Perturbation theory}

Besides the advantages pointed out in section~\ref{advantages} we now face another important advantage of the Brussels School's formalism. Perturbation theory can be applied even in cases where standard quantum mechanics fails.
Those cases are featured by the phenomenon of instability,
which is strongly connected to the sensitivity of a system towards small perturbations. Let the system's Hamiltonian be of the form
\begin{equation}\label{pertu1}
	H_\lambda=H_0+\lambda W\quad,
\end{equation}
where $W$ is the perturbation operator, switched on by the perturbation parameter $\lambda\in[0,1]$, and $H_0$ is the unperturbed part of the Hamiltonian, whose spectral decomposition is known and reads
\begin{equation}\label{spectral1}
	H_0=\sum_n \omega_n|\omega_n\rangle\langle \omega_n|
	+\int_{\sigma_c} d\omega\,\rho_0(\omega)\,\omega|\omega\rangle\langle\omega|\quad.
\end{equation}
To simplify the discussion let the spectrum of $H_0$ be non-degenerate, so $\omega_n\neq \omega_m$ for $n\neq m$. As long as the perturbation strength $\lambda$ is not explicetely needed, it is set to 1 and $H_{\lambda=1}$ is written as $H$.
The spectral decomposition of $H$ is yet unknown and reads in general
\begin{equation}\label{spectral2}
	H=\sum_n E_n|u_n\rangle\langle u_n|
	+\int_{\sigma_c} d\omega\,\rho(\omega)\,\omega|u^\pm(\omega)\rangle\langle u^\pm(\omega)|,
\end{equation}
where the sum is taken over the discrete spectrum $\sigma_d$ and the integration is performed over the continuous spectrum $\sigma_c$ of $H$.

\begin{figure}
	\[\includegraphics[width=0.4\textwidth]{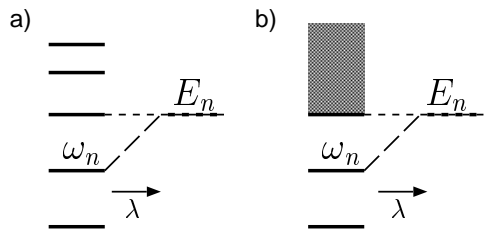}\]
	\vspace*{-0.5cm}\caption{\small a) discrete resonance. b) continuous resonance}\label{fig:shift}
\end{figure}

\begin{figure}[h!]
	\[\includegraphics[width=0.2\textwidth]{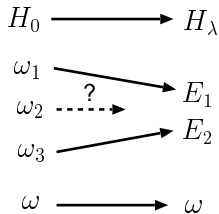}\]
	\vspace*{-0.5cm}\caption{\small In the course of the perturbation the continuous eigenvalues remain unaffected, while the discrete eigenvalues may disappear from the spectrum. In this case the mapping between the discrete spectra is no longer one-to-one and the perturbation series cannot be analytic.}\label{fig:vanish}
\end{figure}

The continuous spectrum $\sigma_c$ is not affected by the perturbation (provided the perturbation operator $W$ is $H_0$-compact, which is assumed here) and there is an identity mapping of perturbed and unperturbed continuous eigenvalues, so in both cases the integration is performed over $\sigma_c$. The discrete case is different. Switching to \emph{Brioullin-Wigner} perturbation theory let us define pairs of orthogonal projection operators
\begin{eqnarray}
	P_n&:=&|\omega_n\rangle\langle \omega_n|\quad\text{and}\\
	Q_n&:=&{\mathbbm1}-P_n\quad.
\end{eqnarray}
Furthermore let the perturbed eigenvector $|u_n\rangle$ of $H$ be not normalized to 1 but rather let $\langle u_n|\omega_n\rangle=1$, such that $P_n|u_n\rangle=|\omega_n\rangle$. 
Applying these projections to the perturbed eigenvalue equation $H|u_n\rangle$ one obtains the \emph{Brioullin-Wigner formulas} for the shifted eigenstate $|u_n\rangle$ and the shifted energy $E_n$:
\begin{eqnarray}
	|u_n\rangle&=&|\omega_n\rangle
	+Q_n\frac\lambda{E_n-H_0}Q_nW|u_n\rangle\quad,\\
	E_n&=&\omega_n+\lambda\langle\omega_n|W|u_n\rangle
\end{eqnarray}
Iteration leads to the corresponding perturbation series
\begin{eqnarray}
	| u_n\rangle&=&\sum_{p=0}^\infty\left\{Q_n\frac\lambda
	{E_n-H_0}
	Q_nW\right\}^p|\omega_n\rangle,\label{briou1}\\
	E_n&=&\omega_n+\lambda\sum_{p=0}^\infty
	\langle\omega_n|W\left\{Q_n\frac\lambda{E_n-H_0}Q_nW
	\right\}^p|\omega_n\rangle.\label{briou2}
\end{eqnarray}
Take a closer look at the above formulas~(\ref{briou1}) and~(\ref{briou2}). 
They are only valid if the expression
\begin{eqnarray}\label{proj1}
	Q_n\frac1{E_n-H_0}Q_n&=&\sum_{m\neq n}
	\frac{|\omega_m\rangle\langle \omega_m|}
	{E_n-\omega_m}\nonumber \\
	&&+\int d\omega\,\rho(\omega)\,
	\frac{|\omega\rangle\langle\omega|}{E_n-\omega}
\end{eqnarray}
is well-defined. Here \emph{resonance} appears as a major obstacle. 
There are two different types of resonance making a perturbation expansion of eigenstates and eigenvalues impossible:
\begin{enumerate}
\item
	\textbf{discrete resonance}: The eigenvalue $\omega_n$ may be shifted in the course of the perturbation to a neighboured eigenvalue of $H_0$ (see fig.~\ref{fig:shift}a). Expression~(\ref{proj1}) is no longer defined and the corresponding perturbation series diverges on account of the summation term.
\item
	\textbf{continuous resonance}: If the eigenvalue $\omega_n$ lies next to the continuous spectrum of $H_0$, it may be shifted there (see fig.~\ref{fig:shift}b). Now the integral term in~(\ref{proj1}) causes a divergence of the perturbation series.
\end{enumerate} 		
In both cases the perturbation expansion is limited to a finite convergence radius of the perturbation parameter $\lambda$ . Outside its range the shifted eigenvalue disappears from the spectrum of $H_\lambda$ and the mapping from the unperturbed to the perturbed spectrum is no longer one-to-one (see fig.~\ref{fig:vanish}). 
Now regard the case of discrete unperturbed eigenvalues \emph{embedded} in the continuous spectrum. Here the convergence radius of $\lambda$ reduces to zero, i.e. the perturbation series are no longer analytic in $\lambda$. An arbitralily small perturbation induces continuous resonance.  As a consequence embedded states get unstable and start to decay. Decaying states cannot be part of the spectral decomposition of $H_\lambda$, they disappear from there and cannot be reconstructed for $\lambda\rightarrow0$, i.e. $H_\lambda\not\rightarrow H_0$. Hence perturbation theory is not aplicable. 
There is a way out: One has to include unstable states, i.e. Gamov vectors, into the spectral decomposition of $H_\lambda$. In terms of perturbation theory the Gamov vectors can be constructed by modified Brioullin-Wigner equations:
\begin{eqnarray}
	|f_k^\pm\rangle&=&|\omega_k\rangle
	+\lim_{E\rightarrow z_k^\pm}Q_k\frac\lambda{E-H_0\pm i\epsilon}Q_kW|f_k^\pm\rangle\quad,\label{briou3}\\
	z_k^\pm&=&\omega_k+\lambda
	\langle\omega_k|W|f_k^\pm\rangle\quad,\label{briou4}
\end{eqnarray}
where $z_k^+=z_k$ are the singularities of $R_+$ on the lower half of the complex plane and $z_k^-=z_k^*$ are those of $R_-$ on the upper half. There is a one-to-one mapping between the disappearing discrete eigenvalues $\omega_k$ and the complex eigenvalues $z_k$.
Iteration leads to the corresponding perurbation series
\begin{eqnarray}
	| f_k^\pm\rangle&=&\lim_{E\rightarrow z_k^\pm}
	\sum_{p=0}^\infty\left\{Q_k\frac\lambda{E-H_0\pm i\epsilon}
	Q_kW\right\}^p|\omega_k\rangle,\label{briou5}\\
	z_k^\pm&=&\omega_k+\lambda
	\langle\omega_k|W|f_k^\pm\rangle\quad.\label{briou6}
\end{eqnarray}
The generalized continuous eigenstates are 
\begin{eqnarray}
	|f^+(\omega)\rangle&=&D_+(\omega)|u^+(\omega)\rangle\\
	\langle f^-(\omega)|&=&\langle u^+(\omega)|\quad,
\end{eqnarray}
with the complex distribution $D_+$ defined in~(\ref{compl_dist}).
The standard continous eigenvectors can be constructed using the
\emph{Lippmann-Schwinger equation}
\begin{equation}
	|u^+(\omega)\rangle=|\omega\rangle+\frac{\lambda}{\omega-H_0+ i\epsilon}W
	|u^+(\omega)\rangle
\end{equation}
or its iteration, the \emph{Born series}:
\begin{equation}
	|u^+(\omega)\rangle=\sum_{p=0}^\infty\left\{\frac{\lambda}
	{\omega-H_0+i\epsilon}
	W\right\}^p|\omega\rangle\quad.
\end{equation}
Take a closer look at the construction of Gamov states in eq.~(\ref{briou3}).
The divergences due to continuous resonance are prevented by introducing a small complex shift in the denominator and analytic continuation to the second Riemann sheet. As a consequence, the energy eigenvalues are shifted to respectively the lower and upper complex plane, inducing exponential decay and Breit-Wigner energy distribution of the corresponding Gamov states. The expressions are still analytic in $\lambda$, so perturbation theory can be applied. This is a major advantage of the Brussels School's formalism.

\section{Example: the Friedrichs model}

The Friedrichs model (K. Friedrichs 1948, see~\cite{friedrichs1948}) is the simplest model involving instability. Here one single discrete state is coupled to a continuum of states by an interaction Hamiltonian $W$. A physical analogon to this model is e.g. the \emph{Wigner-Weisskopf model} (Wigner and Weisskopf 1930, see~\cite{ww}) of a twolevel atom coupled to the vacuum field. Another example is the \emph{Auger effect} of autoionisation (see e.g.~\cite{peres}).

The total Hamiltonian is of the form
\begin{eqnarray}
	H&=&H_0+W\quad,\\
	\text{where }H_0&=&\omega_1|1\rangle\langle1|+\int_0^\infty
		d\omega\,\omega\,|\omega\rangle\langle\omega|\quad,\\
	W&=&\int_0^\infty d\omega\,\Big\{ W(\omega)|\omega\rangle\langle1|
			+W^*(\omega)|1\rangle\langle\omega|\Big\}\quad.
\end{eqnarray} 
Using the \emph{second resolvent identity}
\begin{equation}
	R=R_0+RWR_0\quad,
\end{equation}
with the resolvent $R_0(z)=\frac1{z-H_0}$ of the free Hamiltonian $H_0$ and the resolvent $R(z)=\frac1{z-H}$ of the total Hamiltonian. The resolvent $R(z)$ can be obtained by applying it to the free states $|1\rangle$ and $|\omega\rangle$ and doing some rearrangement:
\begin{eqnarray}
	R(z)&=&\frac1{\eta(z)}\left[|1\rangle+\int_0^\infty d\omega\,\frac{W(\omega)}
	{z-\omega}|\omega\rangle\right]\nonumber \\
	&&\left[\langle1|+\int_0^\infty d\omega\,\frac{W^*(\omega)}
	{z-\omega}\langle\omega|\right]\nonumber \\
	&&+\int_0^\infty d\omega\frac1{z-\omega}|\omega\rangle\langle\omega|\quad.
\end{eqnarray}
The complex function
\begin{equation}
	\eta(z)=z-\omega_1-\int_0^\infty
	d\omega\frac{|W(\omega)|^2}{z-\omega}	
\end{equation}
has analytic extensions $\eta_\pm(\omega)=\eta(\omega\pm i\epsilon)$ from above and below to the real axis. The zeros of $\eta$ on the second Riemann sheet define the complex poles of the resolvent:
\begin{equation}
	\eta_+(z_1)=0,\quad\eta_-(z_1^*)=0\quad.
\end{equation} 
The pole $z_1$ can be obtained by numerically calculating the complex zero of $\eta_+(z)$. In first order one gets
\begin{equation}
	z_1\approx\omega_1+{\mathcal P}\int_0^\infty d\omega\,
	\frac{|W(\omega)|^2}{\omega_1-\omega}-i\pi|W(\omega_1)|^2\quad.
\end{equation}
The Gamov vectors can be obtained using eqns.~(\ref{g_eigenvector1a}) and~(\ref{g_eigenvector1b}):
\begin{eqnarray}
	|f_1^+\rangle&=&\frac1{c_1}\lim_{E\rightarrow z_1}(E-z_1)R_+(E)|1\rangle \\
	&=&\frac1{\sqrt{\eta_+{}'(z_1)}}\left[|1\rangle+\int_0^\infty d\omega\,
	\frac{W(\omega)}{[z_1-\omega]_+}|\omega\rangle\right]\\
	\langle f_1^-|&=&\frac1{c_1}\lim_{E\rightarrow z_1}(E-z_1)\langle1|R_+(E)\\
	&=&\frac1{\sqrt{\eta_+{}'(z_1)}}\left[\langle1|+\int_0^\infty d\omega\,
	\frac{W^*(\omega)}{[z_1-\omega]_+}\langle\omega|\right]\quad,
\end{eqnarray}
where the complex distribution is defined by
\begin{equation}
	\int_0^\infty d\omega\,\frac{\varphi(\omega)}{[z_1-\omega]_+}
	:=\lim_{E\rightarrow z_1}\int_0^\infty d\omega\,
	\frac{\varphi(\omega)}{E-\omega+i\epsilon}\quad.
\end{equation}
The standard outgoing states can be constructed using eq.~(\ref{eigenvector2}):
\begin{eqnarray}
	|u^+(\omega)\rangle&=&\frac1{c(\omega)}
	\frac{i\epsilon}{E-H+i\epsilon}|\omega\rangle\\
	&=&|\omega\rangle+\frac{W^*(\omega)}{\eta_+(\omega)}\times\nonumber \\
	&&\times\left[|1\rangle+\int_0^\infty d\omega'\,\frac{W(\omega')}{\omega-\omega'+i\epsilon}
	|\omega\rangle\right]\quad.
\end{eqnarray}
Using eq.~(\ref{g_eigenvector2}) the generalized continuous eigenvectors read
\begin{eqnarray}
	|f_+(\omega)\rangle&=&D_+(\omega)|u^+(\omega)\rangle\\
	&=&|\omega\rangle+D_+(\omega)\frac{W^*(\omega)}{\eta_+(\omega)}
	\times\nonumber \\
	&&\times\left[|1\rangle+\int_0^\infty d\omega'\,\frac{W(\omega')}{\omega-\omega'+i\epsilon}
	|\omega\rangle\right]\quad,\\
	\langle f^-(\omega)|&=&\langle u^+(\omega)|\\
	&=&\langle\omega|+\frac{W(\omega)}{\eta_-(\omega)}\times\nonumber \\
	&&\times\left[\langle1|+\int_0^\infty d\omega'\,\frac{W^*(\omega')}{\omega-\omega'-i\epsilon}
	\langle\omega|\right]\quad,
\end{eqnarray}
where the complex distribution $D_+$ is defined by~(\ref{compl_dist}) with the curve $\Gamma_+$ leading from $0$ to $\infty$ below $z_1$.

In the pole approximation, i.e. neglecting the background contribution from the continuous eigenvectors, the initial state $|1\rangle$ decays with a decay rate $\Gamma$ given by two times the imaginary part of $z_1$,
\begin{eqnarray}
	P_1(t)&=&|\langle1|U(t)|1\rangle|^2\approx e^{-\Gamma t},\\
	\text{where}\quad\Gamma&=&2*{\mathrm Im}\,z_1\approx2\pi|W(\omega_1)|^2,
\end{eqnarray}
which is in agreement with Fermi's Golden Rule.

\section{Time symmetry}\label{rhs}

\subsection{The Rigged Hilbert Space}\label{rhs2}

When dealing with infinite dimensions, the Hilbert space appears to be too small in the following sense: Position and momentum operator have no Hilbert eigenvectors and Dirac's bra-ket formalism is mathematically not justified. To correct this problem, the Hilbert space must be enlarged to include \emph{improper states} like $\delta$-functions and plane waves. The framework to enable these features is given by the theory of the \emph{Gelfand triplet}, also called the \emph{Rigged Hilbert Space} (RHS). 
\begin{figure}[h!]
	\[\includegraphics[width=0.4\textwidth]{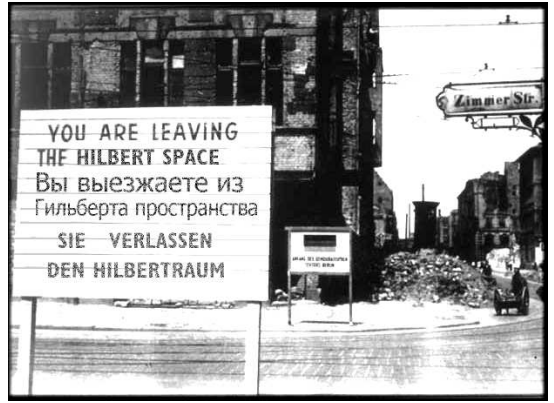}\]
	\vspace*{-0.5cm}\caption{\small To deal with improper states like $\delta$-functions and plane waves and to make Dirac's bra-ket formalism work, one has to leave the Hilbert space.}\label{fig:leaving}
\end{figure}
The concept is the following (comp. e.g.~\cite{bohm1989}).
Take the Hilbert space ${\mathcal H}=L^2$ of squared integrable functions and
find a suitable test space $\Phi\subset{\mathcal H}$ with the properties
\begin{enumerate}
\item
$\Phi$ is a locally convex topological vector space,
\item
$\Phi$ is complete with respect to its own topology,
\item
$\Phi$ is dense in ${\mathcal H}$,
\end{enumerate}
such that the topological dual $\Phi^\dagger$ of $\Phi$, consisting of all linear continuous functionals on $\Phi$, called \emph{distributions}, includes the desired improper states. The Hilbert space itself is his own topological dual, i.e. ${\mathcal H}^\dagger={\mathcal H}$. For any test space $\Phi$ being smaller than $\mathcal H$, the Gelfand triplet
\begin{equation}
	\Phi\subset{\mathcal H}\subset\Phi^\dagger
\end{equation}
makes it possible to extend the bracket $\langle\chi|\varphi\rangle$ to all pairs $\varphi\in\Phi$, $\chi\in\Phi^\dagger$ by
\begin{eqnarray}
	\langle\chi|\varphi\rangle&:=&\chi(\varphi)\\
	\langle\varphi|\chi\rangle&:=&\chi^*(\varphi)\quad.
\end{eqnarray}
On pairs of Hilbert vectors the bracket conicides with the scalar product in ${\mathcal H}$.
The ``kets'' and ``bras'' are now identified by linear functionals on the opposite dual space, i.e.
\begin{eqnarray}
	|\xi\rangle&:=&\langle\cdot|\xi\rangle\\
	\text{and}\quad\langle\xi|&:=&\langle\xi|\cdot\rangle\quad,
\end{eqnarray}
where $\xi\in{\mathcal H},\Phi$ or $\Phi^\dagger$.
An operator $A$ can be extended to the distribution space $\Phi^\dagger$ using
\begin{equation}
	\langle A\chi|\varphi\rangle=\langle\chi|A^\dagger\varphi\rangle\quad.
\end{equation}
Hence $A$ can only be extended to $\Phi^\dagger$ if its adjoint $A^\dagger$ is defined on and does not lead out of the test space $\Phi$:
\begin{equation}\label{extend_op}
	A^\dagger\Phi\subset\Phi\quad.
\end{equation}

The test space of standard quantum mechanics is the Schwartz space ${\mathcal S}$ of rapidly decreasing functions, defined by
\begin{equation}
	{\mathcal S}:=\{f\in C^\infty|\sup_{x}|x^n\partial_x^m f(x)|<\infty\ 
	\forall n,m\}\quad.
\end{equation} 
The Schwartz space is a physically convenient test space, since here position and momentum operator are everywhere defined, they fulfill~(\ref{extend_op}), they are self-adjoint and have their eigenvectors among the \emph{tempered distributions} in ${\mathcal S}^\dagger$. Furthermore, they keep their real spectrum, which makes the RHS $({\mathcal S},L^2,{\mathcal S}^\dagger)$ be called a \emph{tight rigged} Hilbert space. Another popular test space is the space ${\mathcal D}$ of $C^\infty$-functions with compact support. But ${\mathcal D}$ is not tight rigged, since the momentum operator obtains a complex spectrum with complex plane waves $e^{izx}\in{\mathcal D}^\dagger$ as eigenvectors. 

\subsection{Breaking time symmetry}\label{intrinsic}

Now we turn to the RHS of the Brussels School. As already stated  
in section~\ref{deform} the definition of generalized eigenvectors is only possible if the test functions can be analytically continued to whether the lower or the upper half of the complex plane. The retarded right eigenvectors $|f_k^+\rangle$ and $|f^+(E)\rangle$ involve complex distributions continuating the test functions $\varphi^*(E)=\langle\varphi|E\rangle$ to the lower half of the complex plane, denoted by ${\mathbbm C}_-$. There is a class of functions fulfilling this condition: the Hardy class $H_-^2$ from below (for detailed information on Hardy classes see~\cite{duren}). If $\varphi^*(E)$ is in $H_-^2$ then its complex conjugate $\varphi(E)=\langle E|\varphi\rangle$ is in the Hardy class $H_+^2$ from above, whose members can be analytically continuated to the upper half of the complex plane, denoted by ${\mathbbm C}_+$. In detail: The function $\varphi(E)$ is in $H_\pm^2$ if and only if it has an analytic continuation to ${\mathbbm C}_\pm$, such that there is a $C<\infty$ and
\begin{equation}
	\sup_{y>0}\int_{-\infty}^{\infty} dE\,|\varphi(E\pm iy)|^2<C
	\label{hardy_def}\quad.
\end{equation}
Thus, restricted to the real line, Hardy functions are $L^2$-functions.
An equivalent definition can be given using the \emph{Paley-Wiener theorem}: The Hardy spaces $H_\pm^2$ are formed by the inverse Fourier transforms of $L^2$-functions with support on respectively the positive and negative semiaxis ${\mathbbm R}_\pm$:
\begin{equation}
	H_\pm^2={\mathrm F}^{-1}\{L^2({\mathbbm R}_\pm)\}
	\label{paley-wiener}\quad.
\end{equation}
For physical convenience it is necessary, that the test functions are also in the Schwartz class $\mathcal S$. Furthermore, their energy representation can be restricted to the positive semiaxis, since energy is bounded from below by zero. Altogether, the test spaces of the Brussels School are defined by
\begin{equation}
	\Phi_\pm:=(H_\pm^2\cap{\mathcal S})\big|_{{\mathbbm R}_+}
	\label{rhs_brussels}\quad.
\end{equation}
The topological duals $\Phi_\pm^\dagger$ then include the complex distributions used in the generalized eigenvectors. The formalism of the Brussels School can be applied in the RHS
\begin{equation}
	\Phi_\pm\subset{\mathcal H}\subset\Phi_\pm^\dagger\quad.
\end{equation}
Since there are \emph{two} RHS's, one has to investigate how to deal with them.
The retarded unity decomposition~(\ref{unity_ret}) contains the dyadic products $|f_k^+\rangle\langle f_k^-|$. Applying test vectors $\varphi_+$ and $\varphi_-$ to both sides one obtains
\begin{equation}
	\langle\varphi_+|f_k^+\rangle\langle f_k^-|\varphi_-\rangle
	\label{bracket_gen}\quad.
\end{equation}
The above expression only makes sense, if $\varphi_+^*(E)$ and $\varphi_-(E)$ are both in $\Phi_-$, thus $\varphi_+(E)$ must be in $\Phi_+$. Hence the instructions for computing brackets while using the retarded unity decomposition are: On the left hand you have to use $\Phi_+$-vectors and on the right hand $\Phi_-$-vectors. However, since both test spaces are dense in $\mathcal H$, you would not have to care too much, if there was not a significant subtlety: time symmetry is broken. To see this we investigate the time evolution of a test vector $\varphi_-\in\Phi_-$:
\begin{equation}
	\varphi_-^t(E)=e^{-iEt}\varphi_-(E)
	\label{evol-}\quad.
\end{equation}
Since $\varphi(E)$ is in $H_-^2$, there is an analytic continuation $\varphi_-(E-iy)$ to ${\mathbbm C}_-$ and so there is one for its propagated counterpart, given by
\begin{equation}
	\varphi_-^t(E-iy)=e^{-yt}e^{-iEt}\varphi_-(E-iy)\quad.
\end{equation}
Obviously, condition~(\ref{hardy_def}) is violated for $\varphi_-^t$ and $t<0$, thus time evolution leads out of the test space for $t<0$. As a consequence test functions in $\Phi_-$ can only be propagated to the future.
Analogically, test functions in $\Phi_+$ can only be propagated to the past.  
Using condition~(\ref{extend_op}) one finds in addition that the extension of the unitary propagator $U(t)$ to the distribution spaces $\Phi_+^\dagger$ and $\Phi_-^\dagger$ is only possible for respectively $t>0$ and $t<0$. So retarded Gamov vectors $|f_k^+\rangle$ in $\Phi_+^\dagger$ ``propagate'' (as part of the spectral decomposition of Hilbert vectors) only to the future and advanced Gamov vectors $|f_k^-\rangle$ in $\Phi_-^\dagger$ propagate only to the past.
The unitary propagator group ${\mathcal U}$ hence splits into a semigroup ${\mathcal U}_+$ propagating to the future (and applying to vectors in $\Phi_-$ and $\Phi_+^\dagger$) and a second one ${\mathcal U}_-$ propagating to the past (and applying to vectors in $\Phi_+$ and $\Phi_-^\dagger$).
This feature is called by the Brussels School the \emph{intrinsic irreversibility} of quantum mechanics and is identified with a microscopic arrow of time.

\subsection{Time symmetry and Gamov vectors}

Is the special choice of the two RHS's \emph{necessary} for the correct mathematical implementation of Gamov states? If so, one could regard the splitting of the unitary time evolution group into two semigroups as enforced by the mathematical concept. This point of view is taken by \emph{A. Bohm} and \emph{N.L. Harshman} in~\cite{bohm1998} :
\begin{quote}
	``Philosophizing alone would not be enough to take a semigroup instead. To arrive at the semigroup, we start from the empirically desirable properties of Gamov resonance states and let mathematics determine the path.''
\end{quote}
The above mentioned properties are given by
\begin{eqnarray}
	H|f_k^+\rangle&=&(E_R-i\Gamma/2)|f_k^+\rangle\quad,\label{req1}\\
	|f_k^+(t)\rangle&=&e^{-iHt}|f_k^+\rangle
	=e^{-iE_Rt}e^{-i\frac\Gamma2t}|f_k^+\rangle\quad,\label{req2}
\end{eqnarray}
with $E_R,\Gamma>0$.
Starting from these conditions, following the reasoning of section~\ref{intrinsic}, one is led to the two RHS's with broken time symmetry. 
However, the argumentation is not valid if a Rigged Hilbert Space can be found, where Gamov vectors are included and time symmetry is preserved.
Let us start with the space ${\mathcal D}$ of $C^\infty$-functions with compact support. As already mentioned, this space is a perfect test space fulfilling the requirements given in section~\ref{rhs2}. Furthermore we have
${\mathcal D}\subset{\mathcal S}$.
Now regard the space ${\mathcal Z}$ of inversely Fourier transformed ${\mathcal D}$-functions:
\begin{equation}
	{\mathcal Z}:={\mathrm F}^{-1}\{{\mathcal D}\}\quad.
\end{equation}
Since the inverse Fourier transform is a one-to-one mapping on the Schwartz space ${\mathcal S}$, we have
\begin{equation}
	{\mathcal Z}\subset{\mathcal S}\quad.\label{z_schwartz}
\end{equation}
Let $\tilde\varphi(s)$ be a ${\mathcal D}$-function. Its inverse Fourier transform $\varphi(E)\in{\mathcal Z}$ reads
\begin{equation}
	\varphi(E)=\frac1{2\pi}\int_{-\infty}^\infty ds\,e^{iEs}\tilde\varphi(s)
	=\frac1{2\pi}\int_{K} ds\,e^{iEs}\tilde\varphi(s),
\end{equation}
where $K$ is the compact support of $\tilde\varphi(s)$.
The analytic continuation of $\varphi(E)$ to the complex plane exists and is an entire function, given by
\begin{equation}
	\varphi(E+iy)=\frac1{2\pi}\int_{K} ds\,e^{-ys}e^{iEs}\tilde\varphi(s)\quad,
\end{equation}
for all $E,y\in{\mathbbm R}$ (see~\cite{rudin}, theorem 7.22).
Since the inverse Fourier transform is a linear continuous one-to-one mapping, the transformed topology $\tau_{\mathcal D}$ of ${\mathcal D}$, 
\begin{equation}
	\tau_{\mathcal Z}:={\mathrm F}^{-1}\{\tau_{\mathcal D}\}\quad,
\end{equation}
transfers all topological properties from ${\mathcal D}$ to ${\mathcal Z}$. Thus, $\{{\mathcal Z},\tau_{\mathcal Z}\}$ is a topological vector space fulfilling all requirements of a test space in $L^2({\mathbbm R})$. Since ${\mathcal Z}$ contains functions being analytic in the whole complex plane, the topological dual ${\mathcal Z}^\dagger$ contains the retarded Gamov vectors defined by~(\ref{req1}) and~(\ref{req2}). Furthermore, it contains also the \emph{advanced} Gamov vectors $|f_k^-\rangle$ with eigenvalues $z_k^*$ in the upper half of the complex plane, as well as any other generalized eigenvector used in the extended spectral decompositions~(\ref{unity_ret}) and~(\ref{unity_ad}). Because of relation~(\ref{z_schwartz}) all ${\mathcal Z}$-functions are ${\mathcal S}$-functions, so ${\mathcal Z}$ is already a physical convenient test space and standard quantum mechanics is applicable, too. 
Since ${\mathcal D}\not\subset{\mathcal S}({\mathbbm R}_\pm)\not\subset{\mathcal D}$ the space ${\mathcal Z}$ does neither include any of the test spaces $\Phi_\pm$ nor is it part of it, i.e.
\begin{equation}
	{\mathcal Z}\not\subset\Phi_\pm\not\subset{\mathcal Z}\quad,
\end{equation}
but of course ${\mathcal Z}\cap\Phi_\pm\neq\emptyset$.
So ${\mathcal Z}$ is a different test space, whose topological duals include any of the complex distributions needed for a comfortable treatment of decaying systems.

However, there is no intrinsic irreversibility. Let the ${\mathcal D}$-function $\tilde\varphi(s)$ have a compact support bounded by the finite interval $[a,b]$. Its inverse Fourier transform $\varphi(E)$ can be propagated in time yielding $\varphi_t(E)=e^{-iEt}\varphi(E)$, whose Fourier transform,
\begin{equation}
	\tilde\varphi_t(s)=\int_{-\infty}^\infty dE\,e^{-iEs}e^{-iEt}\varphi(E)
	=\tilde\varphi(s+t)\quad,
\end{equation}
has a compact support bounded by the finite interval $[a+t,b+t]$, thus it remains compact for all $t\in{\mathbbm R}$ and $\tilde\varphi_t(s)$ stays in ${\mathcal D}$, so $\varphi_t(E)$ stays in ${\mathcal Z}$. Hence the propagator $U(t)$ does not lead out of the test space for any $t\in{\mathbbm R}$. Consequently, the unitary propagator group ${\mathcal U}=\{U(t)\}$ can be extended to the distribution space ${\mathcal Z}^\dagger$ without any splitting. The support of $\mathcal Z$-functions can be restricted to the positive semiaxis, due to the positivity of energy, and the RHS
\begin{equation}
	\left({\mathcal Z}\subset L^2\subset{\mathcal Z}^\dagger\right)
	\big|_{\mathbbm R_+}
\end{equation}
is thus suitable for the presented concepts.
Any desired unity decomposition including Gamov vectors, retarded or advanced, can equivalently be used. As far as Hilbert states are concerned, no unphysical behaviour is predicted and any calculated physical quantity is numerically identical to the one calculated by standard methods.

\subsection{Physical justification}

Apart from pure mathematics there is a physical reasoning presented by members of the Brussels School who try to derive the special choice of the two Hardy classes from the assumption of \emph{causality} (see e.g. Schulte and Twarock in \cite{doebner} or Antoine, Bohm and Harshman in \cite{bohm1998}, from where the citations below are taken). This is done the following way.

\subsubsection{Argument}

\begin{figure}
	\[{\includegraphics[width=0.4\textwidth]{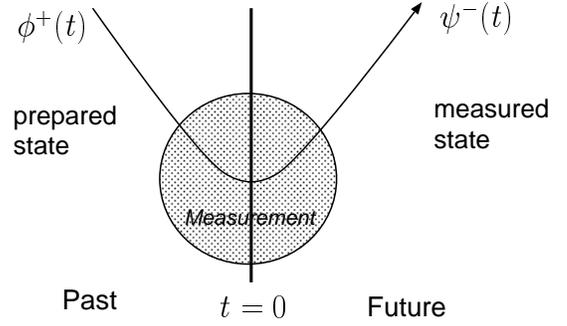}}\]
	\vspace*{-0.5cm}\caption{\small Measurement as a scattering experiment. The prepared state is an incoming state $\phi^+$, the measured state is an outgoing state $\psi^-$ leaving the interaction region where the measurement takes place (grey circle). The survival amplitude of the prepared state is the transition amplitude between incoming and outgoing state.}\label{fig:measurement}
\end{figure}

Starting point is a special assumption of \emph{causality} which is called the ``preparation$\rightarrow$registration arrow  of time'':
\begin{quote}
	``Time translation of the registration apparatus relative to the preparation apparatus makes sense only by an amount of $t\geq0$.''
\end{quote} 
This is translated into the language of quantum mechanics to a so-called ``quantum mechanical arrow of time'':
\begin{quote}
	``A state $\phi^+(t)$ must be prepared before an observable $|\psi^-\rangle\langle\psi^-|=|\psi^-(0)\rangle\langle\psi^-(0)|$ can be measured in that state, i.e. $\phi^+=\phi^+(0)$ must be prepared during $t\leq0$.''
\end{quote}
Interpreting measurement formally as a scattering experiment, the prepared state $\phi^+$ is identified with an incoming state and the measured state $\psi^-$ with an outgoing state (see fig.~\ref{fig:measurement}). Survival probabilities of the initial (i.e. prepared) state are thus calculated by transition probabilities of the incoming and outgoing state:
\begin{eqnarray}
	P(t)&=&|\langle\phi^+|e^{-iHt}|\psi^-\rangle|^2\quad\text{for $t\geq0$,}\\
	P(t)&=&0\quad\text{for $t<0$.}
\end{eqnarray}
Using the energy decomposition this implies
\begin{eqnarray}
	\int dE\,\rho(E)\,e^{-iEt}\phi^+(E)&=&0\quad\text{for $t>0$}\\
	\text{and }\int dE\,\rho(E)\,e^{-iEt}\psi^+(E)&=&0\quad\text{for $t<0$.}
\end{eqnarray}
Using the Paley-Wiener theorem~(\ref{paley-wiener}) one is led to the desired conclusion
\begin{equation}
	\phi^+\in H_+^2,\quad \psi^-\in H_-^2\quad.
\end{equation}
Hence, founding on the given formulation of causality the only correct choice of a Rigged Hilbert Space is the choice of the two RHS's of the Brussels School. In other words: Standard quantum mechanics has not yet found a microscopic arrow of time because it has been incomplete in its axiomatic foundation.

\subsubsection{Critics}

\begin{figure}[h!]
	\[\includegraphics[width=0.3\textwidth]{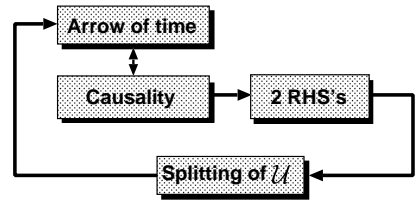}\]
	\vspace*{-0.5cm}\caption{\small A ``causality argument'' equivalent to the postulation of an arrow of time can be used to derive the special choice of two RHS's, which in turn leads to an arrow of time.}\label{fig:circular}
\end{figure}

The above argumentation is correct, but it cannot be taken as a \emph{derivation} of the arrow of time. 
As is obvious even by the choice of words (``before'', ``after'', etc.), the given ``causality'' assumption is already based on the arrow of time, hence it is logically impossible to derive the latter from the former. 
Causality itself does not require an arrow of time, since neither Newton's law nor Schr\"odinger's law contradict causality, although both theories are perfectly time-symmetric. 
Yet, the ``preparation$\rightarrow$registration arrow  of time'' refers explicitely to \emph{open subsystems}, namely the preparation apparatus and the measuring apparatus, which are coupled together to form a composite system.
Though one cannot generally infer from statements about the time evolution of open subsystems to statements about the evoution of closed composite systems, especially not when the subsystems are involved in a measurement process. Hence postulating a ``preparation$\rightarrow$registration arrow  of time'' for \emph{all} systems equals introducing a microscopic arrow of time \emph{by hand} into quantum mechanics. Once the arrow of time has been put into the mathematics, it is not surprising that it will reappear eventually. The situation can be visualized by fig.~\ref{fig:circular}.
Thus, the special choice of the two RHS's must be regarded as being \emph{equivalent to a postulation of a global microscopic arrow of time}. The split RHS's are a {mathematical mirror image} of the empirically motivated splitting between past and future.

\subsection{The arrow of time as a matter of choice}

As already stated in the introduction, it is possible to explain an arrow of time due to time-asymmetric evolution of \emph{macroscopic} observables and due to \emph{local} observations on subsystems. But respecting a \emph{global microscopic arrow of time}, there is a major difference between standard and Rigged Hilbert Space quantum mechanics. In standard quantum mechanics there is no way to implement a global microscopic arrow of time except perhaps by applying brute force and splitting the unitary evolution group axiomatically into two semigroups. In Rigged Hilbert Space quantum mechanics one has the choice of either time-symmetric or time-asymmetric microscopic evolution, due to the choice of the test spaces. Hence, in the RHS formulation it is possible to implement a global microscopic arrow of time in contrast to standard quantum mechanics. So one could argue that nature \emph{obviously} has taken the time-asymmetric choice of two test spaces instead of one. But what does ``obviously'' mean? One might mention our everyday experiences or global time-asymmetric boundary conditions like the big bang or the second law of thermodynamics. Though these experiences and boundary conditions are all related to subsystems or far from being microscopic.
The ``universe'' is a construction from observations on subsystems on a \emph{very} macroscopic scale. Altogether, these time-asymmetric boundary conditions all base on \emph{local macroscopic observations} and hence, as already stated, they do not give rise to a real problem with quantum mechanics, which is a set of microscopic laws. So the choice of global microscopic time-symmetry or time-asymmetry remains merely axiomatic and cannot be derived from other physical postulates.

\section{Summary}

The quantum mechanical formalism, developped by the Brussels School, offers a comfortable treatment of decaying systems, where standard quantum mechanics requires a lot more effort. The generalized spectral decompositions involve Gamov states and can be interpreted in a physically intuitive manner. They directly yield important physical quantities like decay rates and energy level shifts. Perturbation expansions become possible in cases, where standard quantum mechanics fails. 

In contrast to standard quantum mechanics, where the implementation of a microscopic arrow of time is not possible, the Rigged Hilbert Space formalism offers the possibility to incorporate such an arrow by the special choice of two different Gelfand triplets. Due to this choice, the unitary time evolution group splits into two semigroups, each one for respectively the future and the past. However, it is shown that the use of Gamov states does not require this splitting. As a counter example, one single Gelfand Triplet is constructed, where the formalism of the Brussels School can be applied and where time symmetry is preserved. Furthermore, it is pointed out that the physical arguments in favour of the special choice of two Gelfand Triplets are not independent from the arrow of time and thus cannot be used for a derivation of the latter. Hence even within the framework of the Rigged Hilbert Space formalism, the global microscopic arrow of time remains a postulate.

\section{Acknowledgements}

I would like to thank F. Petruccione, H.-P. Breuer, M. Bordemann, and M. Wilkens for stimulating discussions and editorial advice and the referee for valuable comments.
I also thank all other people talking with me about the topic of this paper and contributing in many ways to its completion.

\end{multicols}

\end{document}